\documentclass[prb,showpacs,floatfix,superscriptaddress,amsmath,amssymb,twocolumn]{revtex4-1}
\usepackage{graphicx}
\usepackage{graphics}
\usepackage{dcolumn}
\usepackage{bm}
\usepackage{pstricks}
\usepackage{braket}
\usepackage{amssymb}
\usepackage{hyperref}

\usepackage{float}

\definecolor{lineadecolor}{rgb}{0.35,0.5,0.6}
\definecolor{ddcol}{rgb}{0.8,0.1,0.1}
\definecolor{subsectioncolor}{rgb}{0.1,0.01,0.5}
\definecolor{celeste}{rgb}{0.8,0.87,0.99}

\newcommand{\ba}{\begin{eqnarray}}
\newcommand{\ea}{\end{eqnarray}}
\def\be{\begin{equation}}
\def\ee{\end{equation}}

\begin{document}

\title{Magnon crystals and magnetic phases in a Kagom\'e-stripe antiferromagnet}

\author{S. Acevedo}
\affiliation{IFLP - CONICET. Departamento de F\'isica, Facultad de Ciencias Exactas. Universidad Nacional de La Plata,C.C.\ 67, 1900 La Plata, Argentina.}

\author{ C.A.\ Lamas}
\affiliation{IFLP - CONICET. Departamento de F\'isica, Facultad de Ciencias Exactas. Universidad Nacional de La Plata,C.C.\ 67, 1900 La Plata, Argentina.}

\author{ M.\ Arlego}
\affiliation{IFLP - CONICET. Departamento de F\'isica, Facultad de Ciencias Exactas. Universidad Nacional de La Plata,C.C.\ 67, 1900 La Plata, Argentina.}

\author{ P.\ Pujol}
\affiliation{Laboratoire de Physique Th\`eorique-IRSAMC, CNRS and Universit\`e de Toulouse, UPS, Toulouse, F-31062, France}

\begin{abstract}
In this work we analyze the magnetization properties of an antiferromagnetic Kagom\'e stripe lattice, motivated
by the recent synthesis of materials exhibiting this structure.
By employing a variety of techniques that include numerical methods as Density Matrix Renormalization Group and Monte Carlo simulations, as well as analytical techniques, as perturbative low energy effective models and exact solutions,  we characterize the magnetization process and magnetic phase diagram of a Kagom\'e stripe lattice. The model captures a variety of behaviors present in the two dimensional Kagom\'e lattice, which are described here by analytical models and numerically corroborated. In addition to the characterization of semiclassical intermediate plateaus, it is worth noting the determination of an exact magnon crystal phase which breaks the underlying symmetry of the lattice. This magnon crystal phase generalizes previous findings and according to our knowledge is reported here for the first time.

\end{abstract}

\maketitle

\section{Introduction and Model}

The interplay between geometric frustration and quantum fluctuations enhanced by low dimensionality results in a rich behavior and variety of exotic phases as spin liquids, that despite the theoretical description
\cite{Balents2010,LM_2011,Savary_2016,Knolle2019}
its experimental identification presents great challenges \cite{Wen2019}.\\
A paradigmatic case is the spin-1/2 kagom\'e lattice antiferromagnet, which finds experimental realization in several compounds, such as the Herbertsmithite ZnCu$_3$(OH)$_6$Cl$_2$\cite{K-exp-1}, $\alpha$-vesignieite BaCu$_3$V$_2$O$_8$(OH)$_2$\cite{K-exp-2}, and [NH$_4$]$_2$ [C$_7$H$_{14}$N][V$_7$O$_6$F$_18$]$_5$\cite{K-exp-3}. The spin-1/2 kagom\'e lattice antiferromagnet has been proposed to exhibit a spin liquid ground state, although this aspect has not been fully clarified yet \cite{Hermele2008,Yan1173,Han2012,Fu655,PhysRevB.99.035155}.
\\
Another source of exotic phases are the Bose Einstein condensates (BEC), where a macroscopic number of bosons configure a single particle quantum state \cite{BEC_book}. In antiferromagnetic insulators, the magnetic excitations are usually bosonic magnons, whose interaction with the underlying crystalline lattice can lead to a rich phenomenology, including BEC \cite{Giamarchi2008,RevModPhys.86.563}. \\
The presence of an external magnetic field incorporates an extra degree of freedom that favors the emergence of a variety of behaviors and phases. The simple image of a magnetization curve that grows gradually with the magnetic field until it reaches saturation, in frustrated quantum systems can became considerably more complex.\\
On the one hand, flat regions, called plateaus, can emerge where magnetization remains constant at a certain fraction of saturation, in a range of applied magnetic field \cite{Mila-capitulo-plateaus}. Plateaus can have a classical origin, in the sense that they can be described in terms of relative orientations of classical spins \cite{PhysRevLett.108.057205}. However, there are plateaus that only admit a quantum description, in terms of elemental magnon or spinon excitations \cite{PhysRevLett.82.3168}.\\
Another ingredient is the appearance of jumps in the magnetization curve, due to different mechanisms, such as first-order transitions between classical \cite{PhysRevB.57.11504} or quantum states or BEC of a purely quantum nature \cite{andreas-localized-magnons}.\\
Interacting magnons in a BEC can be localized on certain places of the lattice due to frustration and crystallize through a superfluid-insulator transition, giving rise to a 'magnon crystal' phase \cite{andreas-localized-magnons,Zhitomirsky2005}.\\
Magnon crystal phases are present in a variety of frustrated magnets.
In particular, the spin-1/2 kagom\'e lattice antiferromagnet magnon crystal phases have been predicted below saturation \cite{Nishimoto2013,PhysRevB.88.144416} and found experimentally in the synthetic Cd-kapellasite at very high magnetic fields where the magnons localize on the hexagon of the kagom\'e lattice \cite{Okuma2019}
.\\

\begin{figure}[t]
\begin{center}
 \includegraphics[width=\columnwidth]{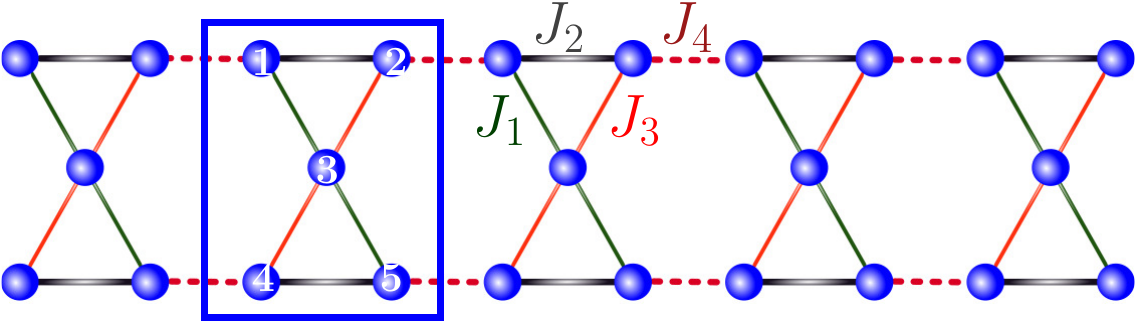}
 \caption{\label{fig:lattice} Geometry of the Kagom\'e stripe lattice considered in this work. The four antiferromagnetic couplings $J_i$, $i=1,\ldots,4$ can be different to model the lattice distortion.
 The unit cell is framed with blue lines and presents geometrical frustration. The white integers 1,\ldots,5 are used to index the sites within the unit cell. }
 \end{center}
\end{figure}

The generality and richness of behaviors described before is also expected in a reduced geometry. This aspect is further enhanced by the recent synthesis of two tellurite-sulfates
A$_2$ Cu$_5$(TeO$_3$)(SO$_4$)$_3$ (OH)$_4$ with A = Na or K \cite{Kagome-exp}. In these compounds, the topological structure of Cu$^{+2}$-ions exhibits a 1D kagom\'e stripe lattice.
The experimental determination of the crystalline structure shows that the kagom\'e stripe is distorted, showing five different Cu-Cu distances,
 as indicated in Fig. 2 (c) of reference \onlinecite{Kagome-exp}. In addition, the study of these compounds suggests an antiferromagnetic behavior and indicates the existence of antiferromagnetic order and some field induced magnetic transitions.\\

Motivated by the mentioned compounds and phenomenology we study the Heisenberg model on the Kagom\'e stripe lattice presented in Fig. \ref{fig:lattice}, in the presence of an external magnetic field.

\begin{equation}
H=\sum_{<i,j>} J_{i,j} \, \vec{S}_{i}\cdot \vec{S}_{j}-h\sum_{i}\vec{S}^{z}_{i} .
\label{hamiltonian}
\end{equation}
We start with a five-spins unit cell in the lattice and four different magnetic couplings $J_{i,j}$ as schematized in Fig. \ref{fig:lattice}.
Note that although the material involves
five different couplings, in this work we consider a space of four couplings.
In this way an extra reflection symmetry is maintained, which simplifies the analysis, without losing the complexity of the unit cell of the material.
This is also justified because it is not intended to describe properly the material.

Throughout this work we will concentrate on different variants and limiting cases of the model to analyze the possible semiclassical and quantum phases that may be present in this system,
and that are of potential interest for the description of the actual materials. In this context we would like to highlight the study by Morita et al \cite{kagome-strip-red}, who analyze the structure of magnetization
curves in the subspace $J_1 = J_3$ (Fig. \ref{fig:lattice}) of our model.

A central aspect of this work is the analysis of the structure of the magnetization curves of the model. In this context the Oshikawa-Yamanaka-Affleck (OYA) theorem\cite{OYA} provides the necessary condition for the presence of magnetization plateaus as
\begin{equation}
NS(1-m)=\hbox{integer},
\label{OYA}
\end{equation}
where $N$ is the number of spins in the ground state unit cell presenting spatial periodicity and $m=\frac{M}{M_{sat}}$ is the normalized magnetization per site.
According to \eqref{OYA}, if the translational symmetry of the lattice is preserved in the ground state ($N=5$), the magnetization curve may have plateaus at $m=1/5$ and $m=3/5$. On the other hand, the emergence of plateaus at different magnetization values is an indication of a spontaneous breaking of the translation symmetry in the ground state.

In this work we will explore both variants of phases that respect or break the underlying symmetry of the lattice, as well as their semiclassical or quantum character. To this end we will use a variety of analytical techniques that will allow us to describe the different emerging plateaus in semiclassical terms or by means of low energy effective models, complemented with numerical methods. The result is a single model with a rich structure of phases, exhibiting semiclassical signatures, as well as truly quantum aspects, as a generalized crystal magnon phase, not reported before.\\

\begin{widetext}

\begin{figure}[t]
\includegraphics[width=0.8\columnwidth]{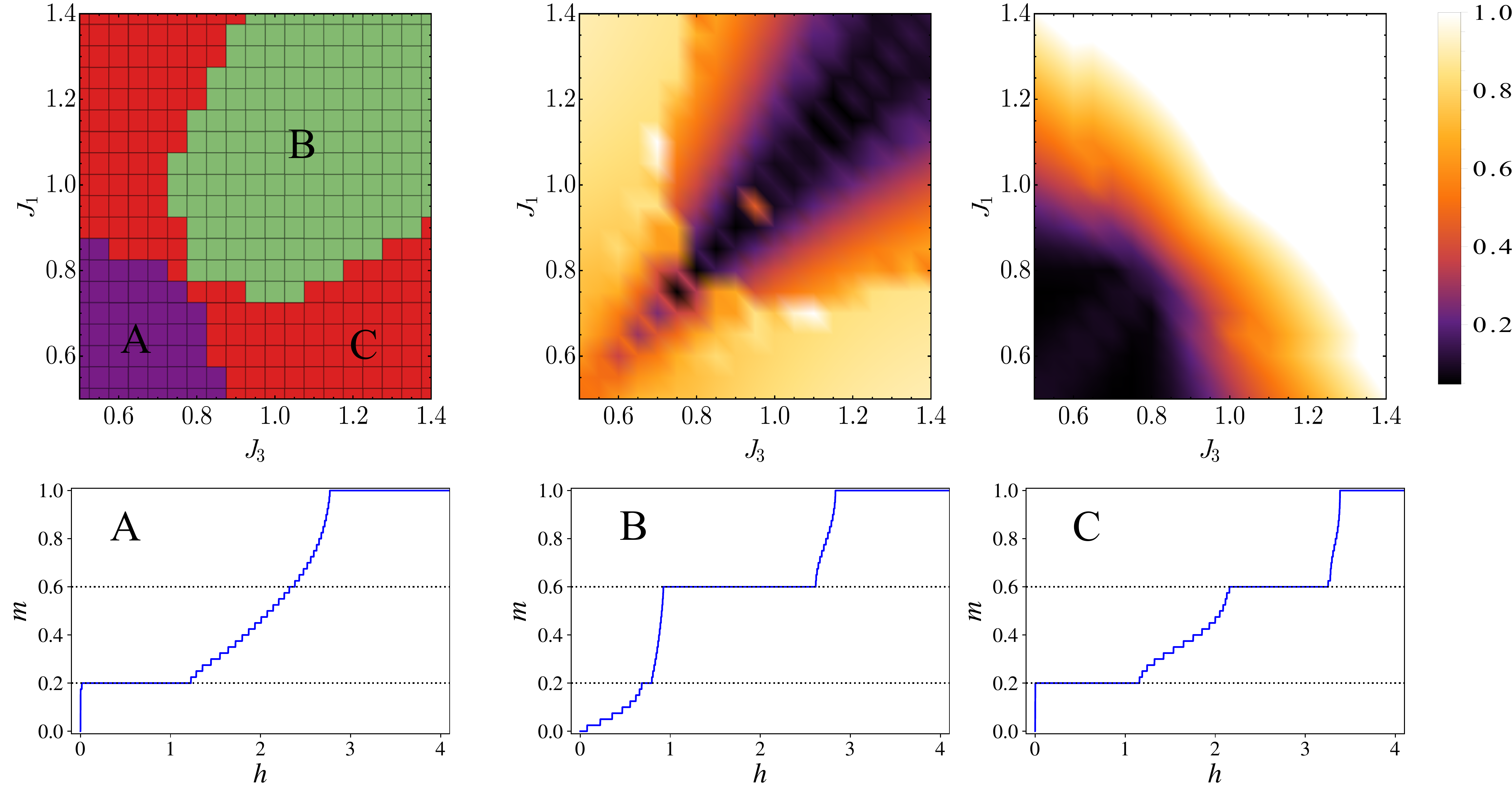}
\caption{Phase diagram of the model in a representative sector of the $J_3-J_1$ plane at $J_2=1$ and $J_4=0.8$,
evaluated by means of DMRG for a stripe with 180 spins (top left panel). The magnetic phases A, B and C are defined by the presence of
magnetization plateaus with $m=1/5$, $m=3/5$ and both, respectively.
Magnetization curves  representative of each phase are depicted in bottom panels. We selected
$(J_3,J_1)=(\frac{7}{15},\frac23)$, $(J_3,J_1)=(\frac{11}{5},2)$, and $(J_3,J_1)=(1,\frac23)$ for A, B and C phases, respectively.
The plateaus widths, in units of $h$, corresponding to $m=1/5$  ($m=3/5$) are showed in top middle (right) panels.}
\label{fig:bigfig}
\end{figure}

\end{widetext}

The paper is organized as follows. In section \ref{sec:classical-plateaus} we study the presence of semiclassical (Ising-like) plateaus in the magnetization curve, and we present the phase diagram in a representative subset of the parameter space. For this we employ Density Matrix Renormalization Group (DMRG) calculations. The main part of this Section concerns to the analysis of the origin of the semiclassical plateaus. This is rooted to the classical Kagom\'e-stripe lattice model and finally to the underlying Ising order of the unit cell. To this end, we study this limiting case explicitly in several Subsections, via Monte Carlo simulations and low energy effective models starting from the strong plaquette (unit cell) limit. The low energy model shows that semiclassical plateaus (in particular $m=3/5$) are perturbatively connected with this limit, completing the analysis of non-breaking lattice symmetry phases of this Section.

In Section \ref{Quantum plateaus} we study phases with spontaneously broken translational symmetry, in which a periodic structure of localized magnons emerges, and in particular at $m=4/5$. For a range of values within the parameters space we find exact ground states of magnon crystal phases \cite{andreas-localized-magnons}.
All these results are complemented with numerical DMRG determination of magnetization curves.

\section{Semiclassical plateaus}
\label{sec:classical-plateaus}

In this Section we analyze the phases of semiclassical plateaus
at $m=1/5$ and $m=3/5$. These phases preserve the lattice translational symmetry,
according to the OYA theorem given by \eqref{OYA}. To determine the extension of the
phases we have evaluated the magnetization as a function of the applied magnetic field for several values
of the couplings, by means of DMRG calculations for large stripes ($200$ spins). The DMRG computations were performed with the open source code ALPS \cite{ALPS}. For the calculations, we kept up to $500$ states throughout the work, which showed to be enough to achieve the required precision. \\
Due to the size of the parameters space, here we illustrate our results in a sector of the $J_3-J_1$ plane at $J_2=1$ and $J_4=0.8$. This subspace captures regions where the plateaus show separately or coexist.

The results are presented in the phase diagram of Fig. \ref{fig:bigfig} (top left), where the
$A,\; B\; \text{or}\; C$ phases
correspond to the presence of a magnetic plateau at $m=1/5$, $m=3/5$ or both, respectively.
At the bottom of Fig. \ref{fig:bigfig} we show a magnetization curve representative of each phase.
Additionally, to evaluate the evolution of the plateaus along the phase diagram,
we determined the plateaus widths for each pair $(J_3,J_1)$. The $m=1/5$ ($m=3/5$) plateaus widths
correspond to the top middle (right) diagram in Fig. \ref{fig:bigfig}.

\subsection{Correlation functions}
\label{sec:correlation functions}
To study the magnetic order associated with the semiclassical plateaus of Fig. \ref{fig:bigfig}, we
computed the $\braket{S^z_1S^z_n}$ correlation function, at $T=0$, using DMRG.

In Fig. \ref{fig:corr-tira} we present the results obtained for the correlation function
vs $n$ (according to the numbering
indicated in Fig. \ref{fig:lattice}). From the top, the first (second) panel corresponds to the magnetic plateau of Fig. \ref{fig:bigfig}
bottom left (middle).
The third and fourth panels of Fig. \ref{fig:corr-tira} correspond to the $m=1/5$ and $m=3/5$ magnetic plateaus of
Fig. \ref{fig:bigfig} bottom right, respectively.\\

The most important aspect to recall is that the correlations structure for all four plateaus analyzed here is in
correspondence with those obtained in the Ising limit, although renormalized by quantum fluctuations.
In the following Subsections (\ref{sec:MC} - \ref{sec:Heff-semiclassical}) we explore this connection
in detail from the perspective of a classical Ising model on the Kagom\'e stripe and from the isolated plaquettes, respectively.

 \subsection{Ising limit of the Kagom\'e stripe}
 \label{sec:Ising-limit}

In order to describe from a classical perspective the magnetic phases analyzed before,
let us first consider the Ising limit of the isolated plaquette (\emph{i.e.} $\vec{S}_{j}=(0,0,S_j^z)$,
and $J_4=0$ in \eqref{hamiltonian}). The Hamiltonian for the plaquette in this case reads
\begin{equation}
 \begin{split}
H_{j}=
 & J_{2}(S^{z}_{j,1}S^{z}_{j,2}+S^{z}_{j,4}S^{z}_{j,5})+
J_{1}S^{z}_{j,3}(S^{z}_{j,1}+S^{z}_{j,5})\\
&+J_{3}S^{z}_{j,3}(S^{z}_{j,2}+S^{z}_{j,4}).
\label{eq:H-ising-plaquette}
\end{split}
\end{equation}
It is possible to identify collinear ground states corresponding to Hamiltonian \ref{eq:H-ising-plaquette}.
Let us consider two different cases (both are 2-degenerate due to spin inversion symmetry), depending of the coupling's ratio:\\

\paragraph*{Case I: $J_{2} > J_{3} > J_{1} \quad (J_2 > J_1 > J_3$)}.
In this case $J_1  (J_3)$ is frustrated and the magnetization of the plaquette is $m=1/5$. The last case is represented in Fig. \ref{fig:casos-Ising-plaquette} (top left).

\paragraph*{Case II: $J_{2}<\{J_{1},J_{3}\}$.}
In this case $J_2$ is frustrated (independently of the relative values of $J_1$ and $J_3$) and the magnetization of the plaquette is $m=3/5$, as represented in Fig. \ref{fig:casos-Ising-plaquette} (top right).\\

These local magnetic structures can be extended to the complete Kagom\'e stripe lattice, where the individual plaquettes are coupled by $J_4>0$.
For the case I, the result of this interaction is a product state of $N$ individual plaquettes in exactly the same state, and the (normalized) magnetization  is still $m=1/5$, as showed in
Fig. \ref{fig:casos-Ising-plaquette} (middle).


\begin{figure}[H]
 \includegraphics[width=0.8\columnwidth]{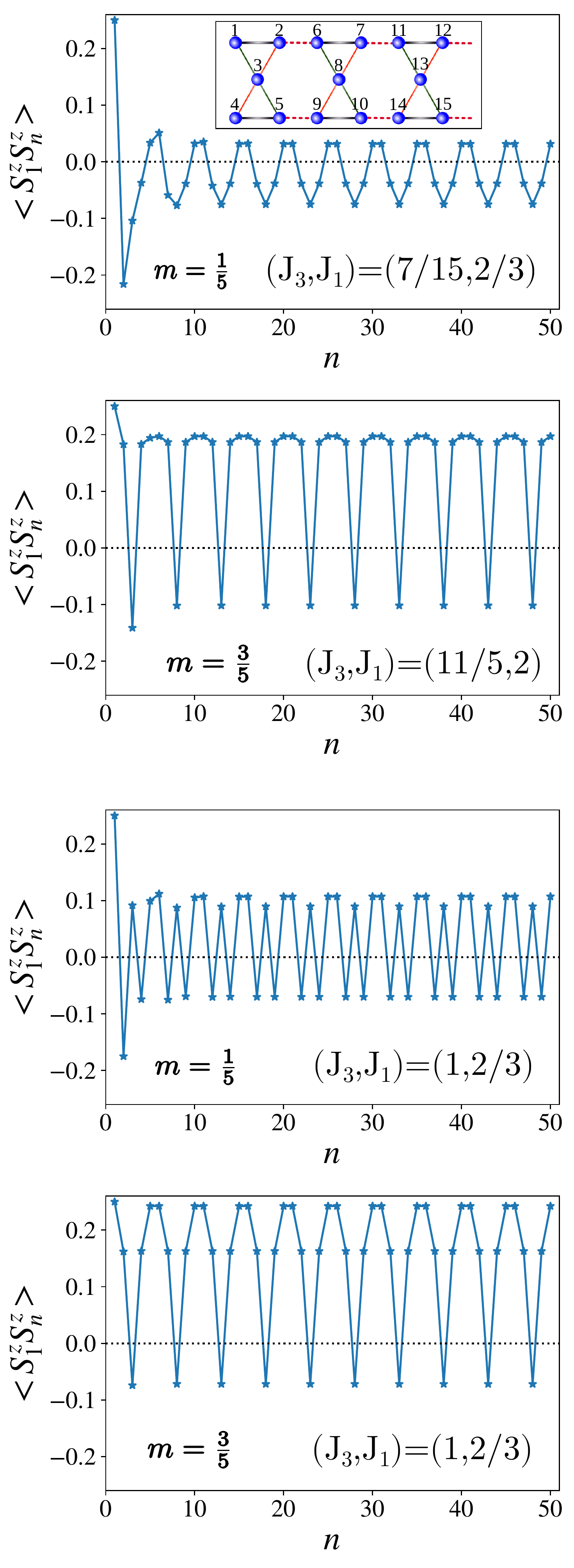}
 \caption{\label{fig:corr-tira} $\braket{S_1^zS_n^z}$ correlation function  on the
 magnetization plateaus representative from each phase showed in Fig. \ref{fig:bigfig} (bottom), calculated by DMRG at T=0, with $J_2=1$.
 The index $n$ follows the indexation from the inset in the first panel of this figure, and $m$ denotes the plateau magnetization.
Note that quantum fluctuations reduce the correlations amplitudes.
However the signature in all four plateaus is in complete agreement with the classical Ising limit, as we
show in Subsections \ref{sec:MC}-\ref{sec:Heff-semiclassical}. }

\end{figure}

In the case II at $h=0$, the $J_4>0$ couples the plaquettes and the stripe has $m=0$. However, at high $h$,
again we can construct a state with $N$ individual plaquettes in the same state, with $m=3/5$,
as we depict in Fig. \ref{fig:casos-Ising-plaquette} (bottom).

\begin{figure}[H]
\begin{center}
 \includegraphics[width=1\columnwidth]{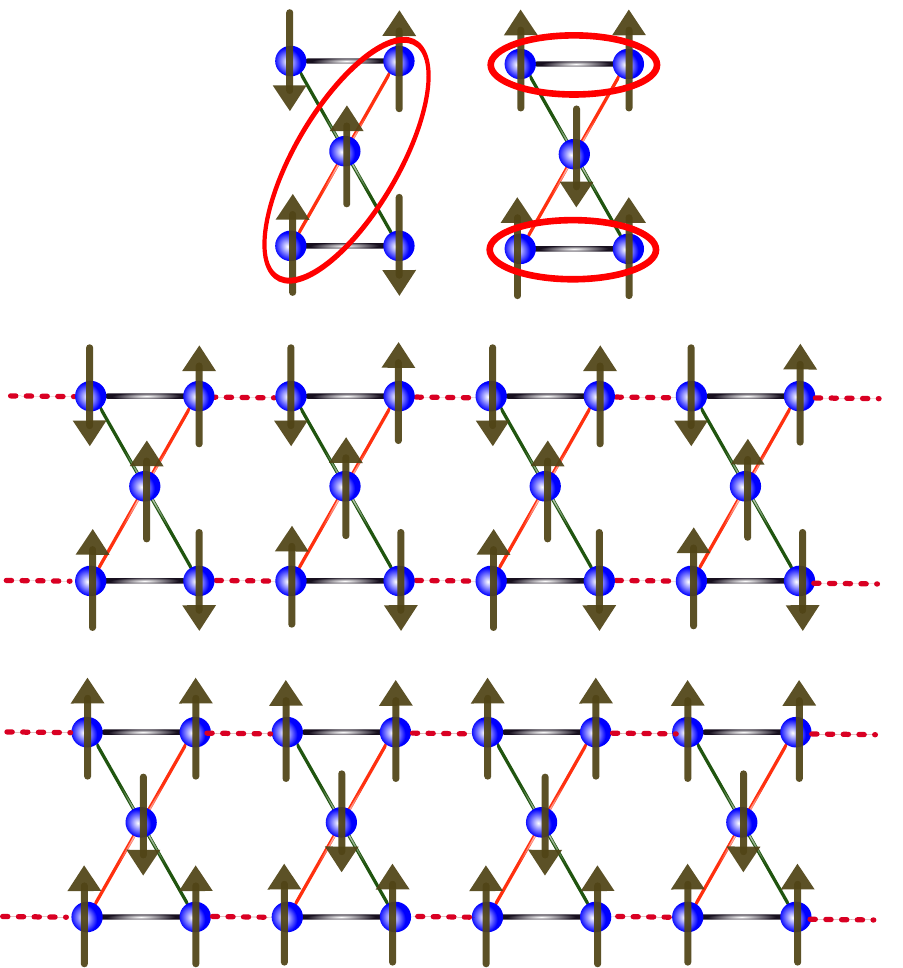}
 \caption{ \label{fig:casos-Ising-plaquette}
 Top left (right): Plaquette Ising ground state for $J_2>J_1>J_3$ }($J_2<\{J_1,J_3\}$) with magnetization $m=1/5$ ($m=3/5$).
 Middle (bottom): Extension of local structures depicted above for finite $J_4>0$ in the Kagom\'e stripe with magnetization $m=1/5$ ($m=3/5$).
 \end{center}
\end{figure}

The above shows that it is possible to construct classical states for the Kagom\'e stripe
(Fig. \ref{fig:casos-Ising-plaquette}) with the same magnetic structure of plateaus at $m=1/5$ and $m=3/5$, obtained via a fully quantum treatment of the model by means of DMRG
(Fig. \ref{fig:bigfig}), which are consistent with Fig. 2 of Morita et al work \cite{kagome-strip-red}. In addition, the correlations calculated by DMRG (\ref{fig:corr-tira}) also show the same structure as the Ising case as it is further investigated in the following subsection.\\

\subsection{Quantum vs thermal fluctuations}
\label{sec:MC}
To compare the role of thermal fluctuations at classical level with zero-temperature quantum effects, we analyzed the finite temperature classical limit of the model. For this we carried out Monte Carlo simulations of the
Ising model, \emph{i.e.} $\vec{S}_{i}=(0,0,S_i^z)$, in \eqref{hamiltonian}, with the Metropolis algorithm
\cite{MC_BOOK}, employing 500 sites, and 1500 independent systems. To prevent the system to stop in a local energy minimum at low $T$, we performed an annealing process, starting with a high temperature state (the system is in
the paramagnetic phase) and then lowering the temperature progressively until no thermal fluctuations are
found.

\begin{figure}[H]
\begin{center}
 \includegraphics[width=\columnwidth]{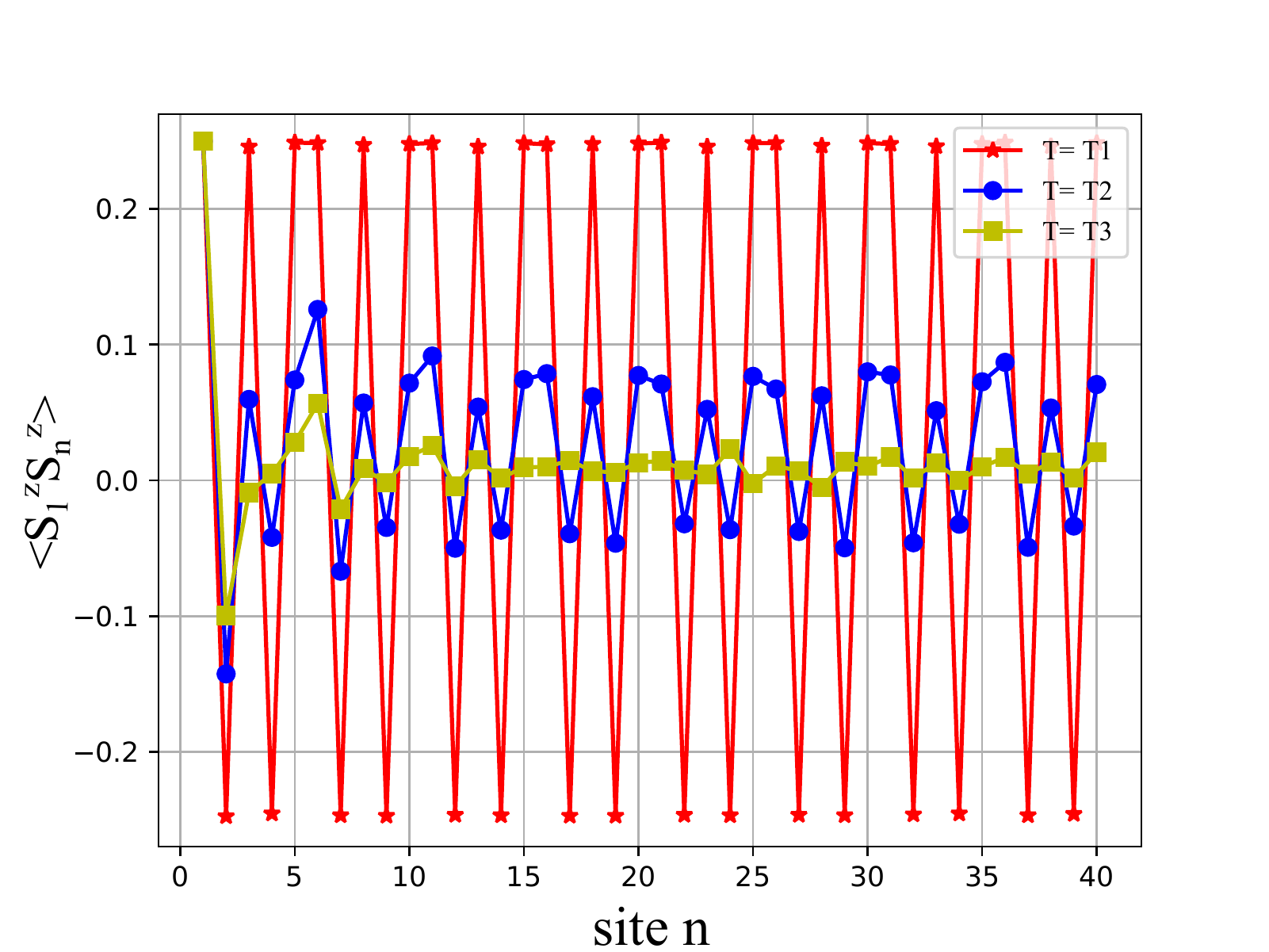}
 \caption{\label{fig:CorrMC} Classical correlation function $\braket{S^z_1S^z_n}$ on the $m=1/5$ magnetization plateau ($h=1.0$) calculated using a Monte Carlo simulation in the Ising limit of the model
 for temperatures $T_1=0.5$ (red), $T_2=1.1$ (blue), and $T_3=1.7$ (yellow), for $J_1=\frac{2}{3}$, $J_2=1$ and $J_3=1$ (same coupling constants as the third panel from top
of Fig. \ref{fig:corr-tira})}. For $T=T_1$ no thermal excitations  are found in the system, indicating that the spins are parallel or antiparallel with each other. For $T=T_3$ the temperature is high
enough for the correlations to rapidly fall off with $n$. For $T=T_2$
the correlations $\braket{S^zS^z}$ obtained by the  \textit{classical}
Monte Carlo  simulation are very similar to those calculated by DMRG using the Heisenberg
model at $T=0$.
 \end{center}
\end{figure}

In Fig. \ref{fig:CorrMC} we present the results of the calculated correlations $\braket{S^z_1 S^z_n}$  for $T_1=0.5$, $T_2=1.1$ and $T_3=1.7$, with $T$ in units of energy.
Note that for $T=T_1$ no thermal excitations  are found in the system, indicating that the spins are parallel or anti-parallel with each other. On the other hand, for $T=T_3$ the temperature is high
enough for the correlations to rapidly fall off with $n$. Finally, for the intermediate $T=T_2$ we highlight
that the correlations $\braket{S_1^zS_n^z}$ obtained by the \textit{classical}
Monte Carlo  simulation are very similar to those calculated by DMRG using the Heisenberg
model at $T=0$. This indicates that the effect of quantum and thermal fluctuations gives an analogous result
in this correlation function.\\

\subsection{Effective model on semiclassical plateaus}
\label{sec:Heff-semiclassical}
\begin{figure}[H]
\begin{center}
 \includegraphics[width=\columnwidth]{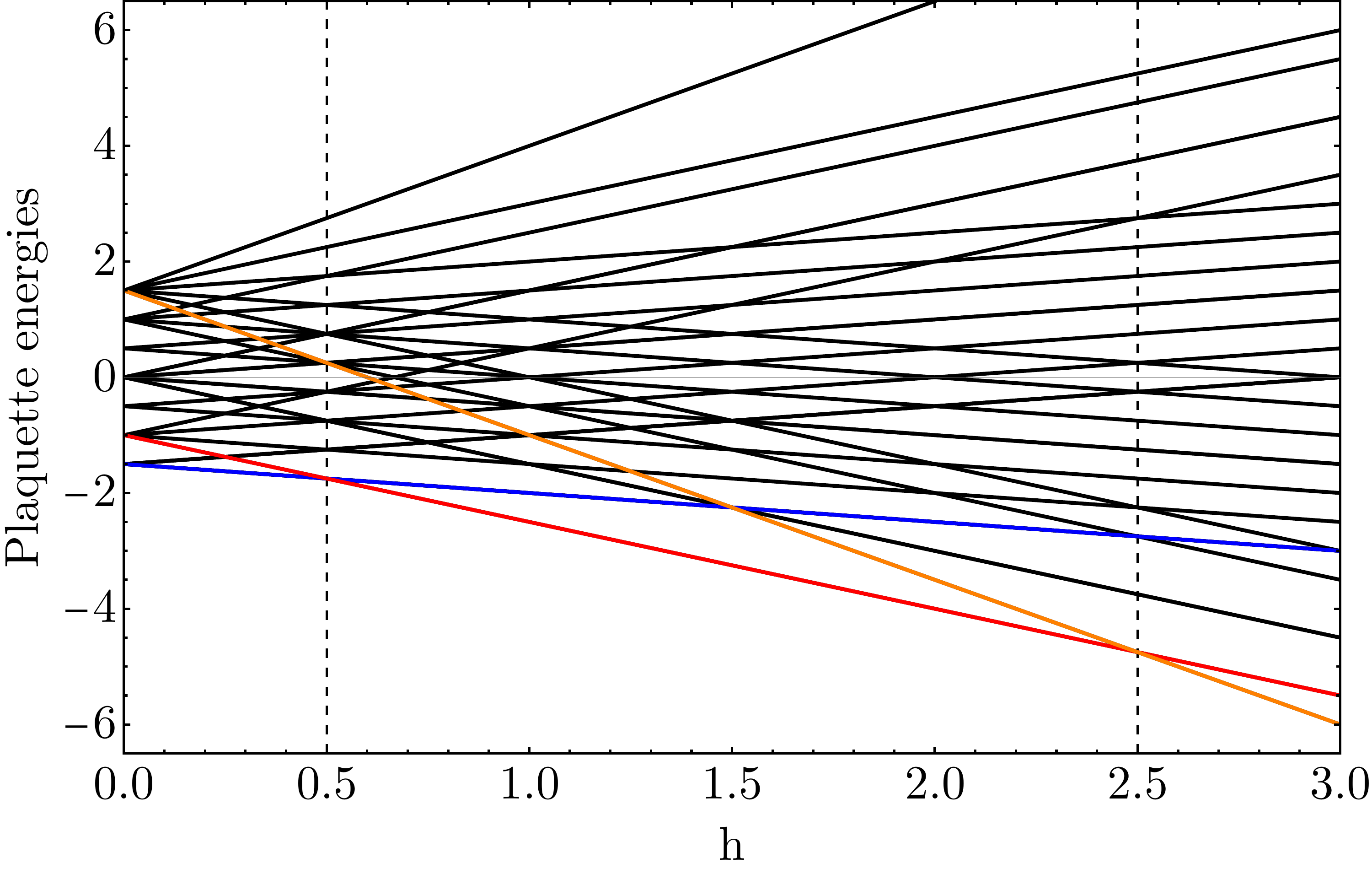}
 \caption{\label{fig:E-plaquette} Energies as functions of magnetic field for the isolated plaquette, for $J_1=J_2=J_3=1$.
Ground states with magnetizations $m=1/5$, $m=3/5$ and $m=1$ are indicated with blue, red and orange lines respectively.
Critical magnetic fields $h_{0}^{(1)}=0.5$ and $h_{0}^{(2)}=2.5$ are highlighted by dashed vertical lines, where the ground state gets three- and two-fold degenerate, respectively.}
 \end{center}
\end{figure}

Here we present an analytical approach for the treatment of semiclassical plateaus, which complements the numerical methods considered before. The method consists in the construction of an effective hamiltonian, based in quantum degenerate perturbation theory \cite{Totsuka1,Mila-ladder-98,Sen-ladder-99,mila-10,4-tube-3,LadderJJd}. In the last part of the work we apply the effective model technique for the case of quantum plateaus that break the translational invariance of the lattice.\\
For the present case, we start by considering a system of isolated plaquettes, \emph{i.e.} $J_4=0$, (see Fig.(\ref{fig:lattice}), whose Hilbert space has dimension $d=2^5$.
For fixed values of the couplings $J_1$, $J_2$ and $J_3$, it is possible to diagonalize numerically
the plaquette Hamiltonian and obtain all the energies as functions of the magnetic field $h$. This is illustrated in
Fig. \ref{fig:E-plaquette} for the homogeneous plaquette case ($J_1=J_2=J_3=1$).\\
Note that for $h>0$ the plaquette has three different ground states corresponding to magnetizations
$m=1/5$ (blue), $m=3/5$ (red) and $m=1$ (orange), depending
on the magnetic field value.
At the critical fields $h_{0}^{(1)}$, $h_{0}^{(2)}$ (corresponding to the dashed vertical lines in the figure),
the ground state gets degenerate. In particular at $h_{0}^{(1)}$ the ground state is three-fold degenerate, although this is a particularity of the homogeneous case where all the couplings are equal.\\

The next step is to consider a  weak coupling between the plaquettes, in particular, at the level crossing.

For $J_4$ finite, we separate the complete hamiltonian in two terms,
\begin{equation}
  H=H_0+H_{int},
\end{equation}
where
\begin{equation}
\begin{split}
&H_0= \sum_n\bigg[ J_1(\vec{S}_{n,1} \cdot \vec{S}_{n,3} + \vec{S}_{n,3} \cdot \vec{S}_{n,5})\\
&+J_2(\vec{S}_{n,1} \cdot \vec{S}_{n,2} + \vec{S}_{n,4} \cdot \vec{S}_{n,5})+
J_3(\vec{S}_{n,2} \cdot \vec{S}_{n,3} + \vec{S}_{n,3} \cdot \vec{S}_{n,4})\\
&-h_0 \sum_{m=1}^5 S^z_{n,m} \bigg]
\end{split}
 \end{equation}
corresponds to the Hamiltonian of a single plaquette,  where $h_0$ is the magnetic field at the energy levels crossing, and

 \begin{equation}
 \begin{split}
  H_{int}= \sum_n\bigg[ J_4(\vec{S}_{n,2} \cdot \vec{S}_{n+1,1} + \vec{S}_{n,5} \cdot \vec{S}_{n+1,4})\\
-(h-h_0)\sum_{m=1}^5 S^z_{n,m} \bigg]
 \end{split}
 \end{equation}
is the plaquettes-interaction term.

Considering $0< J_4, h-h_0 \ll J_i$, $i=1,2,3$; at first order of perturbation theory we have

\begin{equation}
    H^{(1)}=\sum_{ij}\ket{p_i} \braket{p_i|H_{int}|p_j}\bra{p_j},
    \label{1pert}
\end{equation}
where $\ket{p_i}$ are the $2^{N_c}$ degenerated ground states, being $N_c$ the number of unit cell plaquettes.
Computing \eqref{1pert} and expanding the result locally in the $\{ \sigma^0,\frac{\sigma^i}{2} \}$ basis, where $\sigma^0$ is the $2\times 2$ identity matrix and $\sigma^i$ are
the Pauli matrices, one arrives (up to a constant term) at a low energy effective Hamiltonian corresponding to a spin 1/2 anisotropic Heisenberg chain with only nearest-neighbors interactions.
\begin{equation}
 H_{eff}=\sum_n J_{xy} (S^x_n S^x_n+S^y_n S^y_n)+J_{zz} S^z_nS^z_n - \tilde{h} S^z_n,
 \label{Hefff}
\end{equation}
in which the effective couplings $J_{xy}$, $J_{zz}$ and the effective magnetic field $\tilde{h}$
depend on the original couplings $J_i$ and magnetic field $h$. Note that this model is valid for 2-fold degenerate local ground states, which translates into an effective spin-1/2 per site.

For large enough $\tilde{h}$, the ground state of \eqref{Hefff} is the magnon vacuum
$\ket{0}\equiv \ket{\uparrow\uparrow\uparrow\uparrow...}$ (or $\ket{\downarrow\downarrow\downarrow\downarrow...}$).\\
We now compute the 1-magnon dispersion relation

\begin{equation}
  \epsilon_\pm (k) =J_{xy}\, cos(k)-J_{zz} \pm \tilde{h}.
  \label{eq:magnon-dispersion}
\end{equation}
From \eqref{eq:magnon-dispersion} we calculate the  edges of the plateaus around the critical field where the first order expansion is made.
We impose the condition of gap closure, which determines the edge of the plateaus in terms of the magnetic couplings of the effective model.
\begin{equation}
    \tilde{h}= \pm (J_{xy}+ J_{zz}).
    \label{mingapless}
\end{equation}

\begin{figure}[H]
\begin{center}
 \includegraphics[width=0.9\columnwidth]{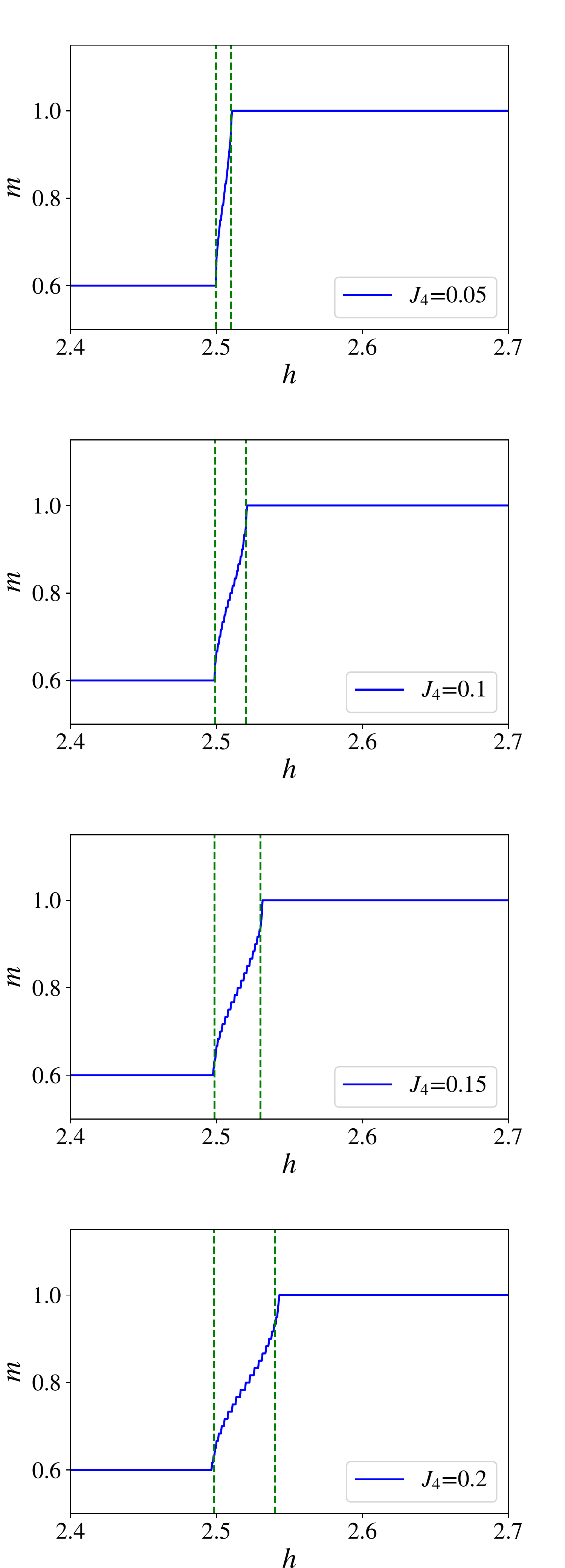}
 \caption{\label{fig:plateau06-Eff}
 Magnetization curves near of saturation for $J_i=1$, $i=1,2,3$ and $J_4=0.05,0.1,0.15,0.2$ from top to bottom, respectively.
 Green dashed lines correspond to the edge of the plateaus predicted by the closure of the gap in the magnon dispersion \eqref{eq:magnon-dispersion}. }
 \end{center}
\end{figure}

In Fig. \ref{fig:plateau06-Eff} we depict four sectors of magnetization curves showing the transition between $m=3/5$ and $m=1$ plateaus.  The green dashed lines mark the plateaus edges as calculated by the low energy effective
Hamiltonian technique, using \eqref{mingapless}, for $J_4=0.05, \; 0.1, \; 0.15, \; 0.2$, and $J_i=1$, $i=1,2,3$. We found a very good agreement in the perturbative regime between analytical and numerical results. \\
The treatment for the case $m=1/5$  is conceptually similar to $m=3/5$ and it has also been studied numerically in other works \cite{kagome-strip-red},
we will not elaborate more on the subject here.

This concludes our identification of \textit{classical} plateaus.
In section \ref{sec:frac-plateaux} the presence of \textit{quantum} plateaus are studied.
These plateaus will not have a classical counterpart correspondence but a purely quantum  mechanical origin which breaks the lattice symmetry.

\section{Quantum plateaus of localized magnons}
\label{Quantum plateaus}
In this section we study phases with spontaneous breaking of the lattice translational symmetry,
in which frustration induces a periodic structure of  localized magnons associated to intermediate (or fractional) plateaus.
Noteworthy, we find an exact magnon crystal ground state with $ m = 4/5 $ of the anisotropic Heisenberg model,
which is a generalization of the state reported by J. Schulenburg et al \cite {andreas-localized-magnons}.

\subsection{A Magnon Crystal Phase in the anisotropic Kagom\'e-stripe}
\label{sec:loc-mag}
\begin{figure}[H]
\begin{center}
 \includegraphics[width=0.8\columnwidth]{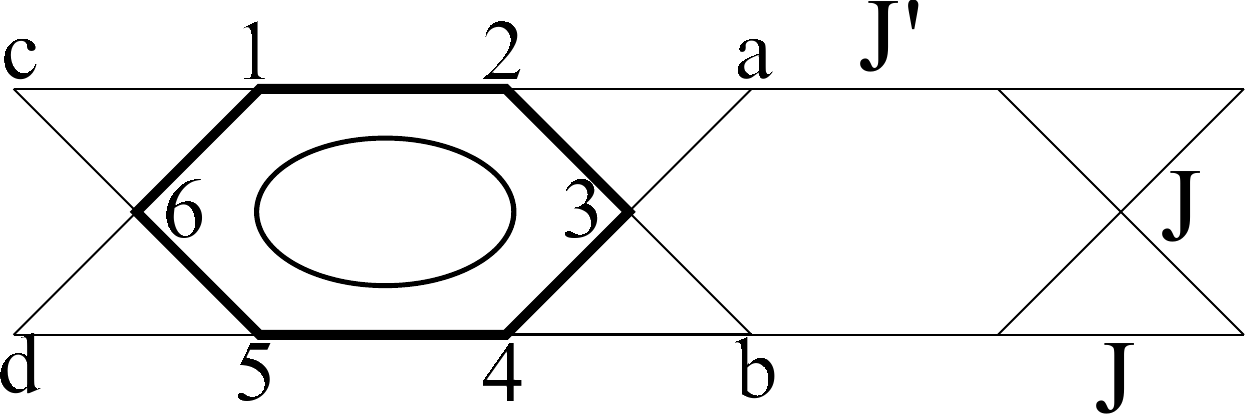}
 \caption{ \label{fig:loc-magnons} Kagom\'e stripe  scheme where the bold hexagon denotes the region L, and the ellipse represents a
 localized magnon. With numbers we index the sites of L, and with letters a,b,c,d we index the sites in interaction with
 the region L. Here we show only two couplings, $J'$ and $J$, both subject to \eqref{eq:loc}. }
 \end{center}
\end{figure}

Let us first consider the anisotropic version of Heisenberg model on the Kagom\'e stripe (\ref{hamiltonian}) in a magnetic field, whose Hamiltonian reads

\begin{equation}
    H= \sum_{<i,j>}J_{ij} \left[
    \Delta S_i^zS_j^z +
    \frac{1}{2}(S_i^+S_j^- + S_i^-S_j^+)\right] - h S^z.
\label{XXZ}
\end{equation}
At high magnetic field the ground state is the fully polarized ferromagnetic state  $\ket{0}\equiv \ket{\uparrow\uparrow\uparrow\uparrow...}$
and the lowest energy excitations can be written in terms of a linear combination of 1-magnon states as
\begin{equation}
 \ket{1}=\sum_l a_l S^-_l \ket{0}.
 \label{eq:state-1}
\end{equation}
Taking a  particular set of coupling values, the magnon dispersion relation may be independent of the momentum $k$
giving rise to a flat band spectrum. This implies that magnon excitations can be localized in a finite region of the stripe. \\
It is possible to construct the exact eigenstate of \eqref{XXZ}, with
localized magnons in the region $L$ represented by the bold hexagon in Fig. \ref{fig:loc-magnons}.

The necessary and sufficient condition for decoupling of the local state from the rest of the system is

\begin{equation}
 \sum_{l \, \epsilon \,L} a_l J_{l\alpha} = 0,
 \label{eq:loc}
\end{equation}
where $J_{l\alpha}$ couples the spins $\vec{S}_l$ and $\vec{S}_\alpha$, with  $l \, \epsilon \, L$ and $\alpha=a,b,c,d$
(see Fig. \ref{fig:loc-magnons}).
It is possible to satisfy \eqref{eq:loc} by taking

\begin{equation}
 a_l=\frac{(-1)^l}{\sqrt{6}}, \; l\, \epsilon \, L,
 \label{eq:a_homogeneos}
\end{equation}

with the indexation of Fig. \ref{fig:loc-magnons},
and the coupling condition

\begin{equation}
  J'=J \frac{2\Delta+1}{\Delta+1},
  \label{CCond}
\end{equation}
where $J'$ couples contiguous two-triangle-cells (the unit cell framed in Fig. \ref{fig:lattice}) and $J$ couples spins inside this unit cell as shown in Fig. \ref{fig:loc-magnons}. This is the exact state reported by J. Schulenburg et al \cite {andreas-localized-magnons}.

The exact ground state of \eqref{XXZ} previous to saturation is the state with $N/2$ localized non interacting magnons
(with $N$ the number of unit-cells) and presents magnetization $m=4/5$. The presence of this exact solution can be observed in the magnetization curve as a jump of $\delta m=1/5$ just below saturation\cite{andreas-localized-magnons}.
Moreover, a fully exactly factorized ground state has necessary short range entanglement entropy. Reciprocally, the property of gapped systems to present short range
entanglement entropy \cite{short_range} gives further indications that, at this point, the system is fully gapped, implying, in particular, the presence of a magnetization plateau. For example,
imagine that one wish to describe the low energy behavior of the system in this point with a field theory description. Among the degrees of freedom of the low energy description,
the magnetic sector is the one related to the presence of a plateau in the magnetization curve: a gap in this sector implies a plateau. On the other hand, in order to obtain a short range entanglement, this
field theory should contain only short ranged or gapped degrees of freedom. This then implies a gap also in the magnetic sector and thus the presence of a magnetization plateau.

\begin{figure}[H]
\begin{center}
 \includegraphics[width=\columnwidth]{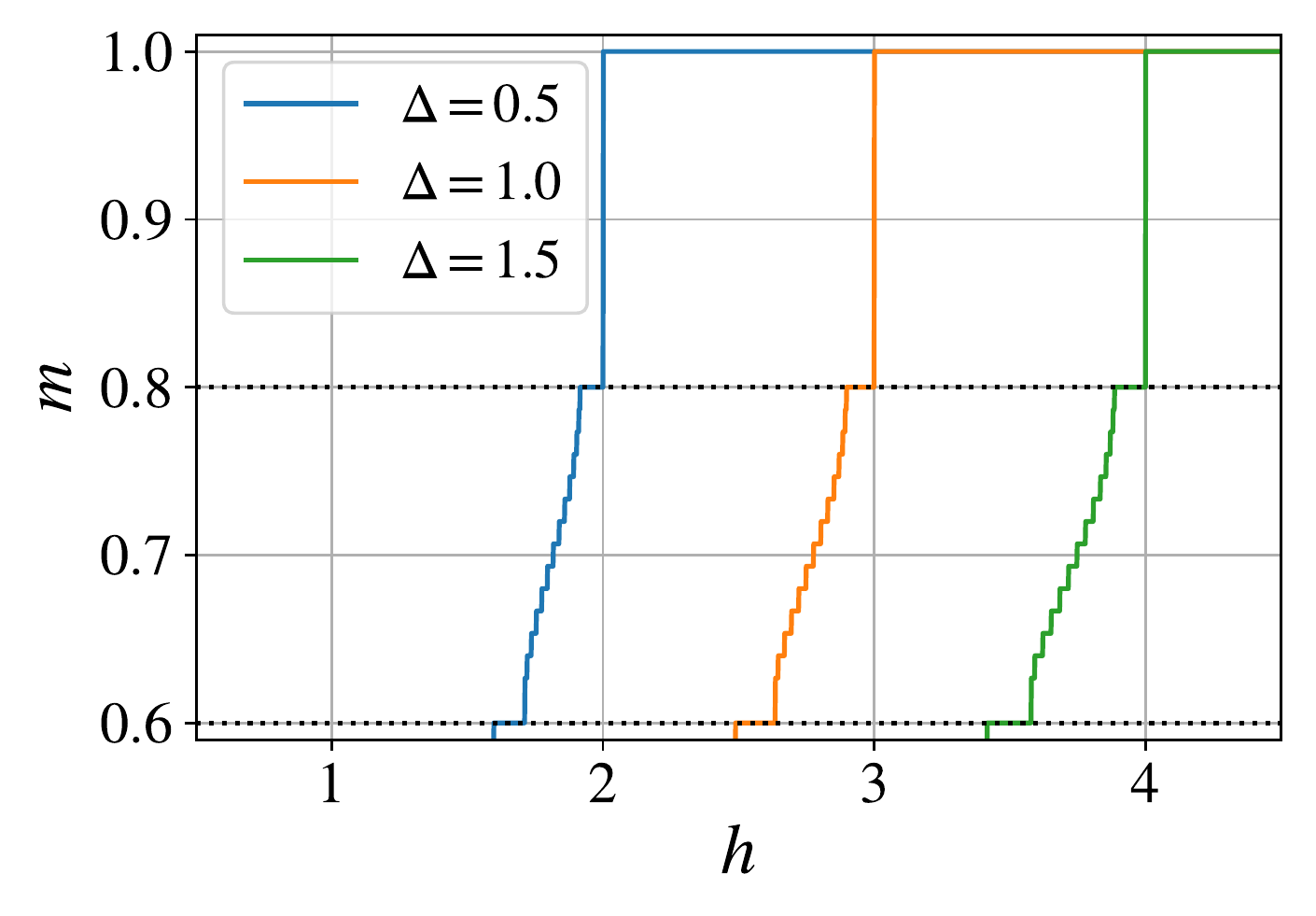}
 \caption{
 \label{fig:jump-magnon}
Macroscopic magnetization jump ($\delta m = 1/5$) to saturation calculated by DMRG for a strip with 150 spins, on the coupling condition \eqref{CCond}
for three different anisotropies, which shows numerically the result of the presence of $N/2$ non-interacting magnons in the stripe as a first excited state.
Additionally, there is a $m=4/5$ plateau.
 }
 \end{center}
\end{figure}

We have constructed the magnetization curve of model (\ref{XXZ})satisfying the coupling condition \eqref{CCond} for several values of anisotropy $\Delta$, by means of DMRG. The results are shown in Fig. \ref{fig:jump-magnon}, where a macroscopic magnetization jump to saturation can be observed.
In addition, a magnetization plateau is present at $m=4/5$. This plateau is consistent with the OYA theorem \ref{OYA} provided that the ground state unit cell contains 10 spins. Therefore, the system breaks spontaneously the
original lattice translation symmetry, doubling the size of the unit cell as is expected for the non-interacting localized magnon state.

\subsection{Effective model on quantum plateaus of localized magnons}

\label{sec:frac-plateaux}
\begin{figure}[H]
\begin{center}
 \includegraphics[width=\columnwidth]{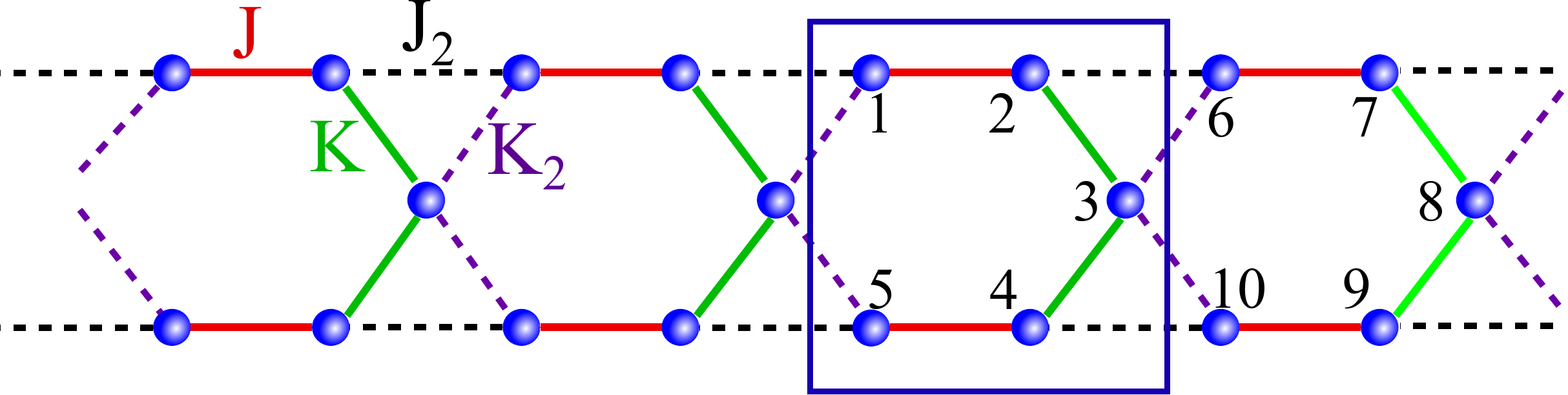}
 \end{center}
 \caption{
 \label{fig:pencil-cell}
 Kagom\'e stripe scheme using the `pencil unit cell' (framed with blue lines). This unit cell is \textit{not} frustrated and composed by five sites, with two couplings, $J$ and $K$, fixed here to 1.5 and 1 respectively. The cells are coupled by $J_2$ and $K_2$. The numbers indicate the sites indexation.
 }
\end{figure}

Here we further study the nature of plateaus with spontaneous breaking of the translational symmetry, in particular $m=4/5$ and the connection with
localized magnons. To this end we constructed a low energy effective hamiltonian via degenerate perturbation theory. First note that according to
(\ref{CCond}), $J'(\Delta=1) = 3/2 J$. This suggests the use of a more convenient unit cell including the strongest coupling $J'$ as shown
in Fig. \ref{fig:pencil-cell}. The unit cells (which we call `\textit{pencil cells}') contain two different couplings $J$ and $K$ and are interconnected
via $J_2$ and $K_2$.

To illustrate the method we fixed $(J,K)=(1.5,1)$, so that for $(J_2,K_2)=(1,1)$ the model satisfies (\ref{CCond}) and the system is
in a exact magnon crystal state, which we call from now on `point-I'.

Our effective model starts from decoupled pencil cells \emph{i.e.} $(J_2,K_2)=(0,0)$, rendering point-I difficult to access pertubatively. However, we will see that the model captures properly the phases with $m = 4/5$, associated with localized magnons.

\begin{figure}[H]
\begin{center}
 \includegraphics[width=0.9\columnwidth]{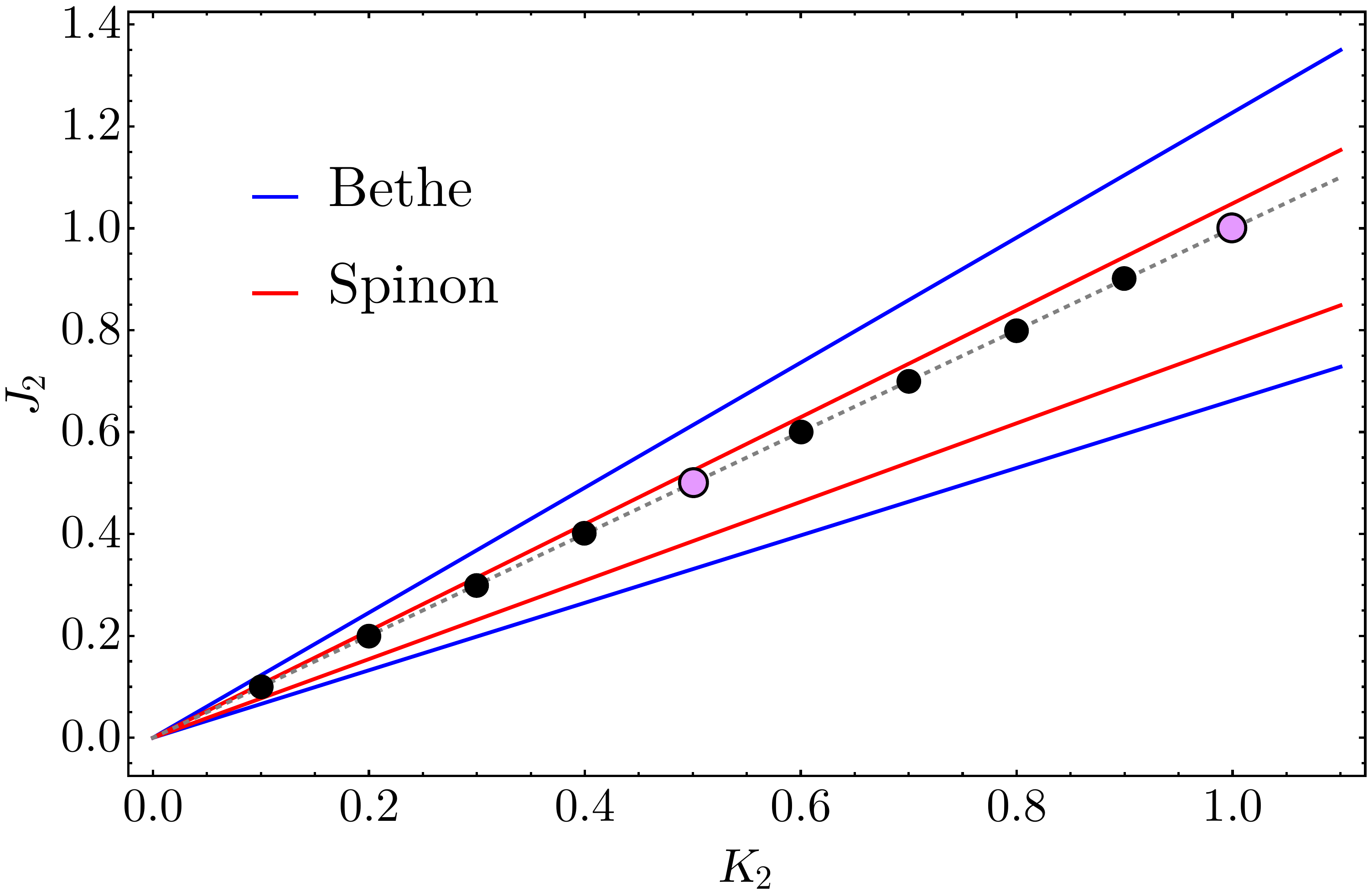}
 \caption{ \label{fig:heff-prediccion-plateau08.pdf}
  Blue lines: the limits of the region where the effective hamiltonian predicts the presence of a fractional m=4/5 plateau using Bethe ansatz.
 Red lines: the corresponding result using the spinon dispersion relation approach on the effective model.  The solid circles indicate the points selected to evaluate, by means of DMRG, the magnetization curves (Fig. \ref{fig:jumps}) and magnon localization (Fig. \ref{fig:correlaciones-xy}). In particular, pink circles denote the presence of a exact solution, crystal magnon phase, of localized magnons.
 }
 \end{center}
\end{figure}

To construct the effective model we start by separating the hamiltonian into $H=H_0+H_{int}$, where

\begin{equation}
\begin{split}
&H_0= \sum_n\bigg[ J(\vec{S}_{n,1} \cdot \vec{S}_{n,2} + \vec{S}_{n,5} \cdot \vec{S}_{n,4})\\
&+K(\vec{S}_{n,2} \cdot \vec{S}_{n,3} + \vec{S}_{n,3} \cdot \vec{S}_{n,4}) -h_0 \sum_{m=1}^5 S^z_{n,m} \bigg],
\end{split}
 \end{equation}
in which $h_0$ is, as in subsection \ref{sec:Heff-semiclassical}, the magnetic field where the isolated pencil-plaquette ground state gets
degenerated due to the level crossing, and
 \begin{equation}
 \begin{split}
  &H_{int}= \sum_n\bigg[ J_2(\vec{S}_{n,2} \cdot \vec{S}_{n+1,1} + \vec{S}_{n,4} \cdot \vec{S}_{n+1,5})+\\
  &K_2(\vec{S}_{n,3} \cdot \vec{S}_{n+1,1} + \vec{S}_{n,3} \cdot \vec{S}_{n+1,5}) -(h-h_0)\sum_{m=1}^5 S^z_{n,m} \bigg].
 \end{split}
 \end{equation}
By performing first order perturbation theory as before, we get a low energy effective Hamiltonian that predicts a region in couplings space where a fractional plateau at $m=4/5$ emerges. We proceeded in two ways, as showed in
Fig. \ref{fig:heff-prediccion-plateau08.pdf}. In blue we plot the solutions for $\Delta=1$ in the effective model which, according to
Bethe Ansatz \cite{bethe}, indicates that the effective chain does not pass through the N\'eel phase\cite{LadderJJd}, and consequently the
Kagom\'e stripe does not have a fractional plateau.

In red, we plot the solutions for a gapless spinon dispersion relation over the N\'eel state where both critical magnetic fields (plateau edges) are equal.
Both regions contain the exact solution $(J_2,K_2)=(1,1)$ (point-I), represented by a pink circle in Fig. \ref{fig:heff-prediccion-plateau08.pdf}, although the model is truly valid only in the $J_2,K_2 \ll J,K$ limit.

\begin{figure}[H]
\begin{center}
 \includegraphics[width=\columnwidth]{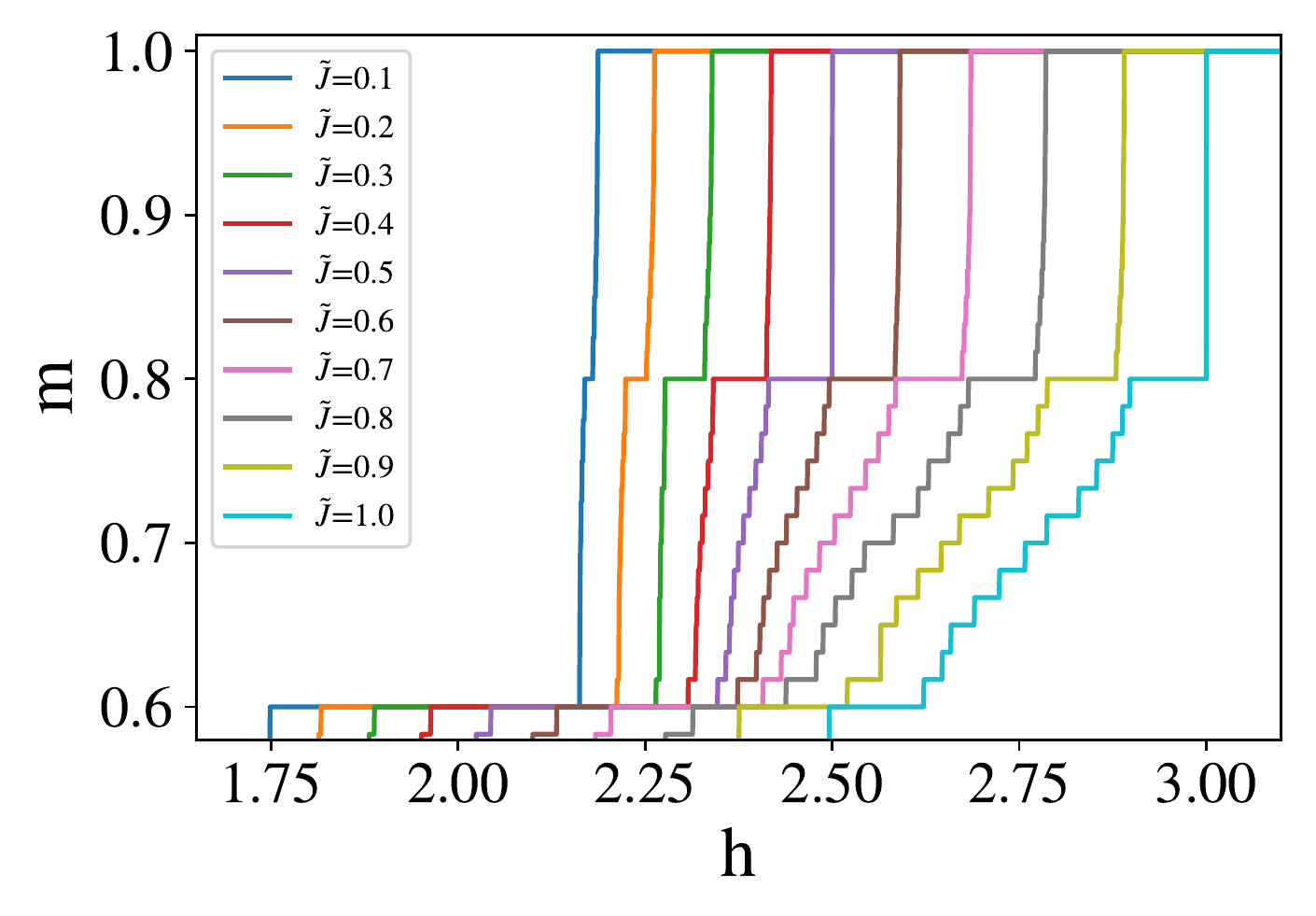}
 \caption{ \label{fig:jumps} Partial magnetization curves computed by means of DMRG for $J=3/2$, $K=1$ and $J_2=K_2\equiv \tilde{J}=0.1,0.2,...1$
 using 24 pencil unit cells. Although there is a large slope transition in all the cases, for $\tilde{J}=0.5$ there is a clean jump from $m=4/5$ to saturation,
 as in the $\tilde{J}=1$ case (point-I). This is a signature of a \textit{second} exact solution of localized magnons which we named `point-II'.
 }
 \end{center}
\end{figure}

To analyze the evolution of the fractional $m=4/5$ plateau from the perturbative regime until reaching point-I, we evaluated numerically the magnetization curves by DMRG in the line $J_2=K_2 \equiv \tilde{J}  $, for $\tilde{J}=0.1,0.2,..., 1$ (points represented by circles in Fig. \ref{fig:heff-prediccion-plateau08.pdf}), with $\Delta=1$. \\
The results are presented in Fig. \ref{fig:jumps}. As it can be observed in all cases there is a transition to saturation with a large slope, together with an increase in the m = 4/5 plateau width. In particular, for $\tilde{J}=0.5$ there is an abrupt jump of $\delta m=1/5$, identical to the magnetization jump observed for point-I ($\tilde{J}=1$). Motivated by this result,  we proposed a state of the form \eqref{eq:state-1} and found a \textit{second} exact solution of localized magnons, that we named `point-II', represented by a pink circle at $J_2=K_2=0.5$ in Fig. \ref{fig:heff-prediccion-plateau08.pdf}.

To analyze the transition between hexagon to pencil cell localized magnons, we numerically evaluated the $\braket{S^+_iS^-_j}$ correlation functions, by means of DMRG.
The results are presented in Fig. \ref{fig:correlaciones-xy}, where we show $\braket{S^+_iS^-_j}$ for $J_2=K_2\equiv \tilde{J}=0.1,0.2,...,1$ (corresponding to the ten dots in Fig. \ref{fig:heff-prediccion-plateau08.pdf}).
As it can observed, in the homogeneous case where $\tilde{J}=1$  the fluctuations are localized on the hexagons as predicted. Note that this particular case is also depicted in Fig. 5 (a) of
Morita et al work \cite{kagome-strip-red}. On the other hand, as $\tilde{J}$ decreases, the fluctuations localize on pencil cells progressively.

\subsection{A generalized magnon crystal phase}

\begin{figure}[H]
\begin{center}
 \includegraphics[width=0.9\columnwidth]{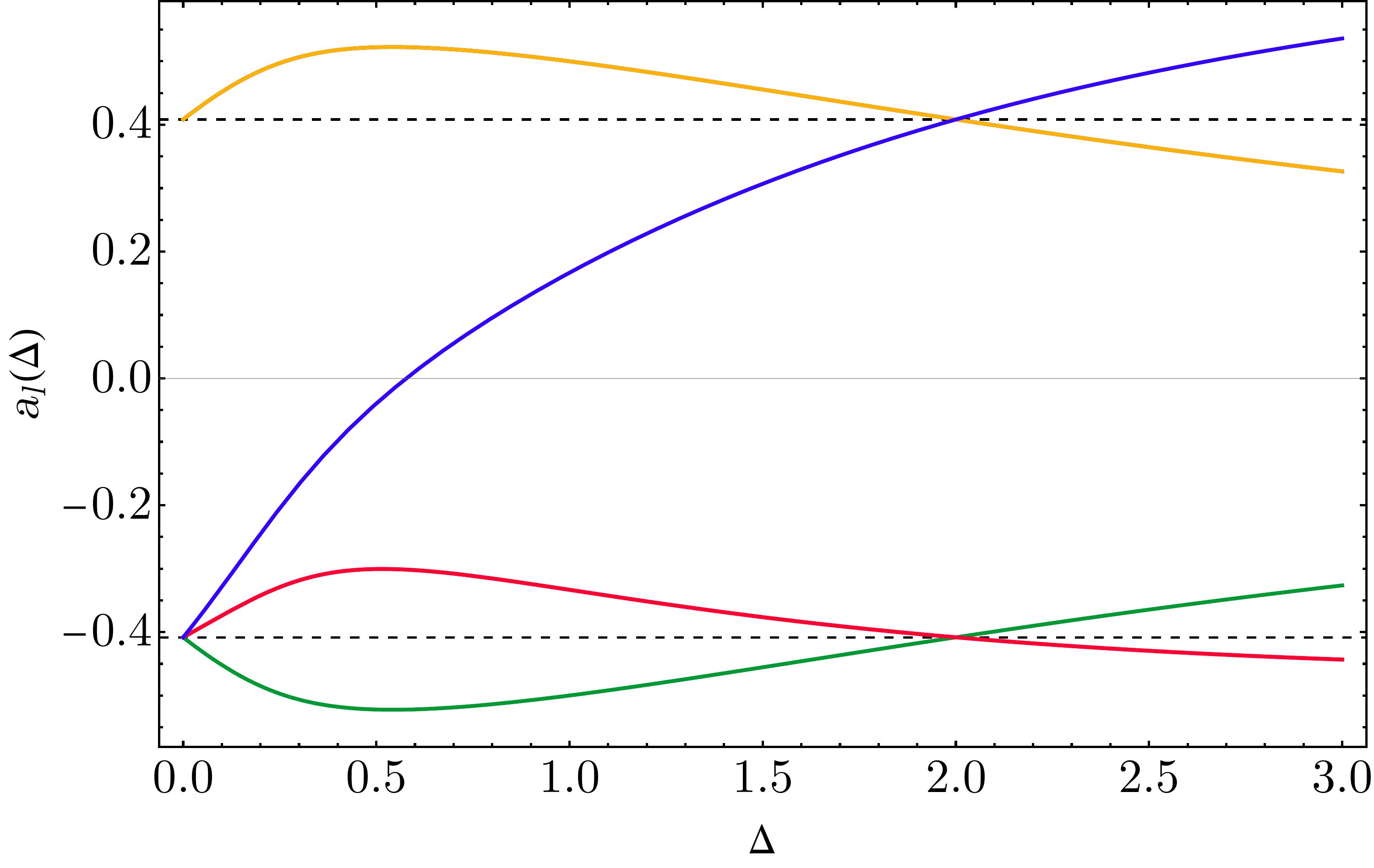}
 \caption{ \label{fig:coef-punto-ii-D} Amplitudes $a_1=a_5$ (pink), $a_2=a_4$ (yellow), $a_3$ (green) and $a_6$ (violet) as functions of
 the anisotropy $\Delta$, for the generalization of the magnon crystal phase of point-II. With dashed horizontal lines, the solution  \eqref{eq:a_homogeneos} (exact ground state in
 point-I).
 Both states are equal at $\Delta=2$.
 }
 \end{center}
\end{figure}

\begin{figure}[t]
\begin{center}
 \includegraphics[width=0.9\columnwidth]{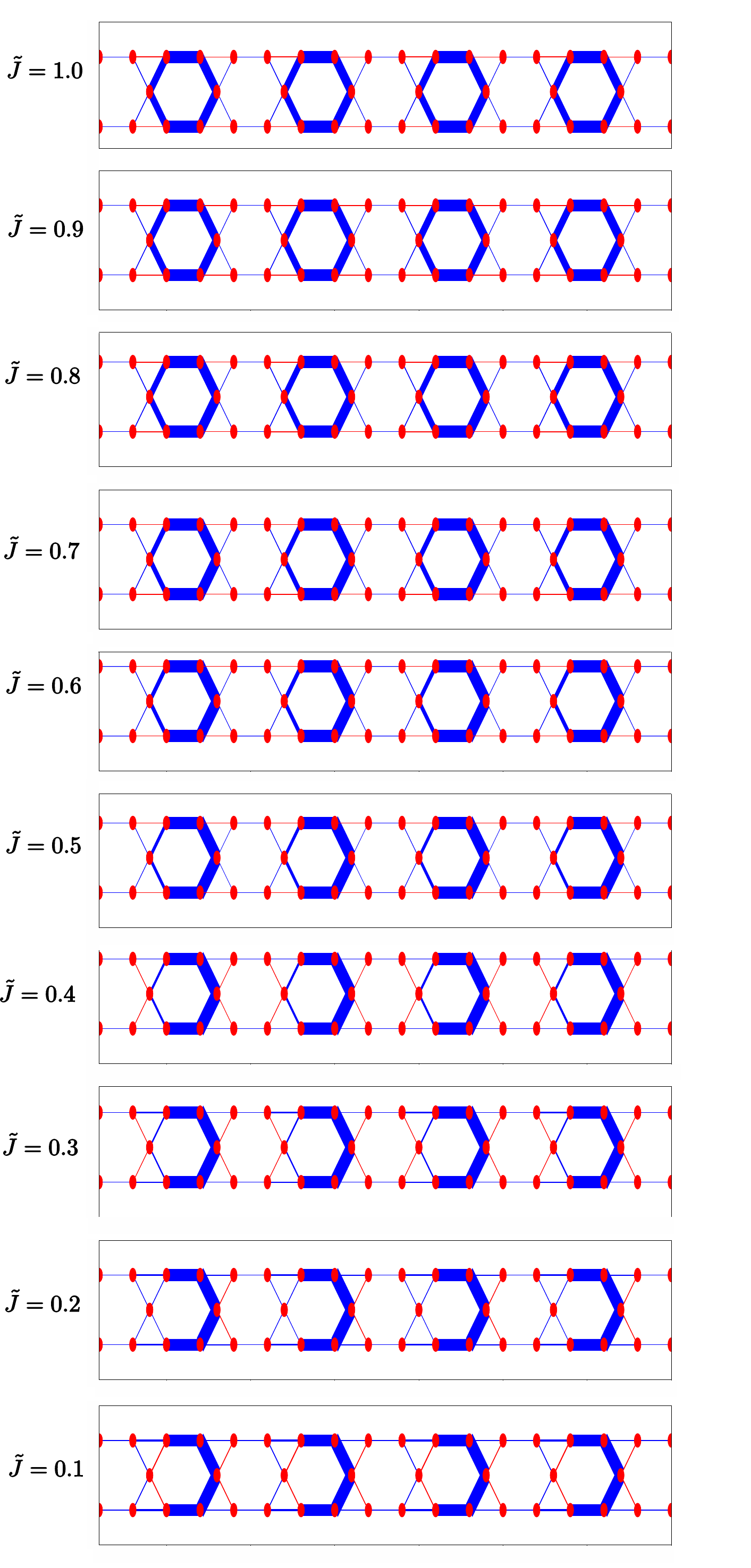}
 \caption{ \label{fig:correlaciones-xy} Transition between hexagon to pencil cell localized magnons.
  Line widths indicate the $\braket{S^+_iS^-_j}$ correlation, for $\Delta=1$, $J=3/2$, $K=1$ and  $K_2=J_2\equiv \tilde{J}=0.1,0.2,...,1$, computed by DMRG.  Blue (red) lines correspond to negative (positive) correlations. The fluctuations are located on hexagons or pencil cells for $\tilde{J}$ near the unity or zero, respectively.}
 \end{center}
\end{figure}

To generalize the solution in point-II to the anisotropic case  ($\Delta \neq 1$) we proposed a coupling set\\
$ \{J=\frac{2\Delta+1}{\Delta+1}K, K=1, J_2(\Delta)=K_2(\Delta) \}$ (following de notation from
Fig. \ref{fig:pencil-cell}) and we found the solution

\begin{equation}
    \frac{K_2(\Delta)}{K}=\frac{\sqrt{12 \Delta ^3 ( \Delta+1)+1}-(2 \Delta +1)}{2\Delta (\Delta +1) },
 \label{eq:sol_delta_punto-II}
\end{equation}
together with a set of couplings $a_l(\Delta)$, $l\, \epsilon \, L$, plotted in Fig. \ref{fig:coef-punto-ii-D}.

We highlight that for $\Delta=2$ the coefficients $a_l$ are exactly \eqref{eq:a_homogeneos}, as showed in
Fig. \ref{fig:coef-punto-ii-D}; while the couplings are $\{J=5/3, K=1, J_2=K_2=1 \}$. In addition, $K_2(\Delta=2)$ and $K$ have
the same value as in the point-I case, while $J$ is different. Finally let us note that,
for $\Delta \, \epsilon \, (0, 1/\sqrt{3})$, \eqref{eq:sol_delta_punto-II} is negative, indicating ferromagnetic couplings.

\section{Conclusions}

In the present paper we studied the magnetization properties of an antiferromagnetic Kagom\'e stripe lattice.
We constructed a magnetic phase diagram which shows three magnetic phases with the presence of a $m=1/5$, $m=3/5$ magnetic plateaus,
or both simultaneously. These plateaus are classical in the sense that they can be understood in terms of the Ising limit,  by studying
the magnetization curves and the $\braket{S^z_iS^z_j}$ correlation function as well as comparing quantum density matrix renormalization group (DMRG)
calculations with classical Monte Carlo simulations in the Ising ground state for different coupling configurations.

We calculated the plateaus edges by means of the low energy effective Hamiltonian technique in the strong plaquettes limit.
The same technique proved to be remarkably useful in predicting the presence of a fractional $m=4/5$ quantum plateau; a plateau that cannot be explained in the Ising Limit.
This plateau is bounded to the presence of a localized magnon phase, as can be seen by computing the
$\braket{S^+_iS^-_j}$ correlation function with DMRG. Furthermore, we found an exact ground state with $m=4/5$ (just before saturation, due to a magnetization jump) of the anisotropic Heisenberg  Hamiltonian,
that provides a generalization of the state found by  J. Schulenburg et al\cite{andreas-localized-magnons}.
This gives another example of a an exactly factorized magnon crystal ground state which finds its origin in the strong frustrating nature of the Hamiltonian \cite{plat2015selection,PI2}.\\

From a more general point of view, magnon crystals are known to be present in a wide variety of one and two dimensional frustrated systems \cite{andreas-localized-magnons}.
The hallmark of these system is a magnon flat band producing an exactly factorized ground state of
localized magnons which is purely quantum mechanical. The magnetic phase diagram of the model studied
here has the richness of having both, this kind of factorized quantum state as well as magnetization
classical (Ising like) plateaus. In particular this phenomenology is also present in the fully 2D kagom\'e model, indicating that some essential aspects of the system transcend dimensionality.
This has been an additional motivation to study the kagom\'e stripe model, which also provides a more accessible numerical treatment.\\
Finally, the richness of this system makes it an ideal laboratory for studying the behavior of such different gaped states in the presence of perturbations like transverse field or Dzyaloshinskii-Moriya interactions.

\section*{Acknowledgments}
We acknowledge useful discussions with M. Matera.
C. A. Lamas is supported by ANPCyT (PICT 2013-0009)

\bibliography{referencias}


\end{document}